\documentclass[twocolumn,showpacs,preprintnumbers,amsmath,amssymb]{revtex4}
\usepackage{graphicx,epsfig}
\usepackage{dcolumn}
\usepackage{bm}

\def\2te{2{\theta}}
\def\chic#1{{\scriptscriptstyle #1}}

\def\'#1{\ifx#1i\accent19\i\else\accent19#1\fi}

\def\8{\infty}

\def\ref#1{$^{#1}$}

\def\'#1{\ifx#1i\accent19\i\else\accent19#1\fi}

\def\8{\infty}

\def\ref#1{$^{#1}$}
\begin{document}
\title{Neutrinos and Nucleosynthesis in Supernova
\footnote{Presented at Proceeedings of the {\it Mexican School 
of Astrophysics} ({\bf EMA}), Guanajuato, Mexico, July 31 - August 7, 
2002. Final version to appear in the {\bf Proceedings of 
IX Mexican Workshop on Particles and Fields} {\it ``Physics Beyond the 
Standard Model''}, Colima Col. M\'exico, November 17-22, 2003.} }
\author{U. Solis}
 \email{solis@nuclecu.unam.mx}
\author{J. C. D'Olivo}
 \email{dolivo@nuclecu.unam.mx}
\affiliation{Instituto de Ciencias Nucleares,\\
             Departamento de F{\'i}sica de Altas Energ{\'i}as,\\
             Universidad Nacional Aut\'onoma de M\'exico (ICN-UNAM).\\
             Apartado Postal 70-543, 04510 M\'exico, D.F., M\'exico.}
\author{Luis G. Cabral-Rosetti}
 \email{luis@nucleares.unam.mx ; lgcabral@ciidet.edu.mx}
\affiliation{Departamento de Posgrado,\\
Centro Interdisciplinario de Investigaci\'on y 
Docencia en Educaci\'on T\'ecnica (CIIDET),\\
Av. Universidad 282 Pte., Col. Centro, A. Postal 752, C. P. 76000,\\
Santiago de Queretaro, Qro., M\'exico.}
\begin{abstract}
\noindent
The type II supernova is considered as a candidate site for the production of 
heavy elements. The nucleosynthesis occurs in an intense neutrino flux, we 
calculate the electron fraction in this environment.
\end{abstract}
\pacs{12345678990}
\maketitle

\section{Nucleosynthesis in Supernova}

A star lives a luminous life by burning $H$ into successively heavier 
elements. However, as the Fe group nuclei near mass number $A=56$ are 
most tightly bound, no more nuclear binding energy can be released to 
power the star by burning Fe. Therefore, heavy elements beyond Fe have 
to be made by process other than normal stellar burning. One such 
process is the rapid neutron capture process, or the r-process. 

One starts with some nuclei and lots of neutrons, the nuclei rapidly 
capture these neutrons to make very neutron-rich unstable progenitor 
nuclei. After neutron capture stops, the progenitor nuclei successively 
beta-decay towards stability and become the r-process nuclei observed 
in nature. This process is responsible for approximately half the 
natural abundance of nuclei with mass number $A>100$ 
\cite{Quian}.

There is as yet no consensus for the site or sites of r-process 
nucleosynthesis. The high neutron densities $10^{20}$ $cm^3$ and 
temperatures of $10^9$ K associated with r-process suggest astrophysical 
sites as core-collapse type II or Ib supernovae. The most plausible 
environment yet proposed is the neutrino-heated ejecta from the nascent 
neutron star. Close to the neutron star, the temperature is several 
Mev and the atmosphere is essentially dissociated into neutrons and 
protons. As the neutrinos emitted by the neutron star free-stream through 
this atmosphere, some of the $\nu_e$ and  ${\overline{\nu}_e}$ are 
captured by the nucleons and their energy is deposited in the atmosphere. 
The atmosphere is heated and, as a result, it expands away from the 
neutron star and eventually develops into a mass outflow, a 
{\it neutrino-driven wind} \cite{Caldwell}. After the shock 
wave has propagated out to several hundred kilometers, condition 
behind the shock at 100 to 200 km are suitable for neutrino heating. 
The neutrino heating blows a hot bubble above the proton-neutron star.

\section{Neutrino-Nucleon Interaction}

Neutrinos and antineutrinos of all three flavors are emitted by the 
neutron star producided in a supernova. The individual neutrino species 
has approximately the same luminosity but very diferent average energy. 
As the neutrinos diffuse out of the neutron star, they thermally decouple 
from the neutron star matter at different radii due to the diference 
in their ability to exchange energy with such matter. Neutrinos species 
of all flavors have identical neutral current interactions but, due to 
energy threshold effects, the $\nu_{\mu}$, $\nu_{\tau}$, and their 
antiparticles lack the charged current capture reactions analogous to 
\begin{eqnarray}
\label{e.1}
\nu_e + n \rightarrow p + e \\
\ \ \ \ \ \ \ \ \ \ n + e^+ \rightarrow p + {\overline{\nu}_e}.
\end{eqnarray}
%

The result is that $\nu_{\tau}$, $\nu_{\mu}$, and their antiparticles, 
have indentical spectra and decouple at a higher density, and thus 
temperature, than the electron neutrinos and antineutrinos \cite{Nunokawa1}.

We require the neutron-to-nuclei ratio $R>100$ to effect a good r-process 
yield for heavy nuclear species, but the models with conventional equations 
of state for nuclear matter all give smaller values of $R$. In all models 
$R$ is  determined by the net neutron-proton ratio; the entropy-per-baryon 
in the ejecta, $s$; and the dynamic expansion time scale, $t_{dyn}$ . 
The neutron-to-proton ratio is $n/p =  Y_e^{-1}-1$, where $Y_e$ is the 
number of electrons per baryon. The r-process is only possible when 
$Y_e<0.5$ at freeze-out from nuclear statistical equilibrium. The value 
of $Y_e$ in the region above the neutrinosphere is determined by the 
interactions in Eq. (1) and (2). We can write the rate of change of 
$Y_e$ with time as
\begin{equation}
\frac{dY_e}{dt} = \lambda_1 - \lambda_2 Y_e,
\end{equation}
\noindent
where $\lambda_1= \lambda_{n \nu_e}+ \lambda_{e^+ n}$ and 
$\lambda_2 = \lambda_1 + \lambda_{n \bar{\nu_e}} +\lambda_{e p}$, are 
the rates in (1) and (2), (see Ref\cite{Nunokawa2}).

\begin{figure*}
\includegraphics[height=.25\textheight]{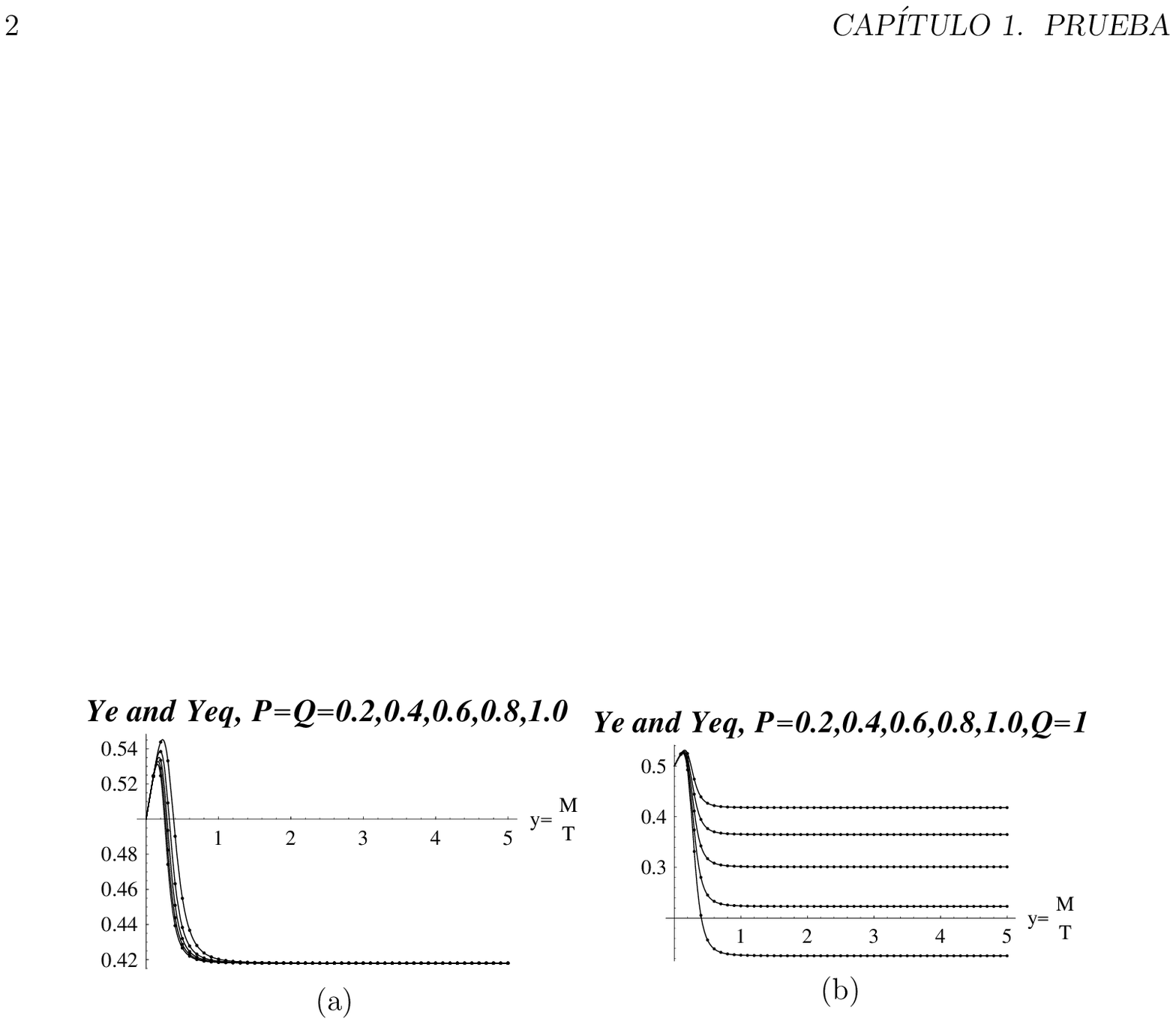}
\caption{In (a) the electron fraction $Y_e$ (doted line) and $Y_{eq}$ 
(solid line) is plotted against $y=(m_n - m_p)/T$ with 
$P = Q = 0.2, 0.4, 0.6, 0.8, 1.0$. In (b) the same but assuming that 
no exist antineutrino oscillations, i. e. $Q = 1$.}
\label{vertex}
\end{figure*}
\section{Electron Fraction}
The general solution to the above equation is given by 
\begin{equation}
Y_e = \frac{\lambda_1}{\lambda_2} - \int_0^t I(t,t^{'}) 
\left[\frac{d}{dt^{'}}
\left( \frac{\lambda_1(t^{'})}{\lambda_2(t^{'})} \right) \right] dt^{'} , 
\end{equation}
\noindent
with the integrating factor given by 
\begin{equation}
I(t,t^{'})= Exp \left({-\int_{t^{'}}^t \lambda_2(t{''})dt{''}}\right)
\end{equation}
\noindent
and  $\lambda_1/ \lambda_2 \equiv Y_{eq}$. The functional form for 
the weak-interaction rates are given by: 
\begin{equation}
\begin{array}{c}
\displaystyle
\lambda_{\nu n} = \frac{B\, L_{\nu}}{r^2 < E_{\nu} >}
\\[0.5cm]
\displaystyle
\times \frac{1}{T^3_{\nu} F_2(0)}
\int^{\infty}_0 (1 - f_e)\, E_{e}^2\, P\, f_{\nu}\,  d E_{\nu}\, ,
\end{array}
\end{equation} 
\begin{equation}
\begin{array}{c}
\displaystyle
\lambda_{{\bar \nu} p} = \frac{B\, L_{{\bar \nu}}}{r^2 < E_{{\bar \nu}} >}
\\[0.5cm]
\displaystyle
\times \frac{1}{T^3_{{\bar \nu}} F_2(0)}
\int^{\infty}_{{\Delta}m} (1 - f_e)\, E_{e}^2\, {\bar P}\, f_{{\bar \nu}}\,  
d E_{{\bar \nu}}\, ,
\end{array}
\end{equation} 
\begin{equation}
\lambda_{e^{-} p} = A 
\int^{\infty}_{{\Delta}m} (1 - P\, f_{\nu})\, 
E_{\nu}^2\, f_{e^{-}}\,  d E_{e^{-}}\, ,
\end{equation} 
\begin{equation}
\lambda_{e^{+} n} = A 
\int^{\infty}_{0} (1 - {\bar P}\, f_{{\bar \nu}})\, 
E_{\bar \nu}^2\, f_{e^{+}}\,  d E_{e^{+}}\, ,
\end{equation} 

\noindent
where $B = 9.6 \times 10^{-44} cm^2/MeV^2$, 
$L_{\nu} = 1 \times 10^{51}\, erg/seg$ and 
$L_{{\bar \nu}} = 1.3 \times 10^{51}\, erg/seg$ are the 
neutrino luminosity, $< E_{\nu} > = 3.15\, T_{\nu}$, $F_2(0) = 1.8$, 
$f = E^2/(e^{E/T} + 1)$ is the distribution function of the neutrinos or
electrons and positrons. $(1 - f_e)$ and $(1 - f_{\nu})$ are the Pauli 
blocking factor and  $r$ is evaluated in $10\, km.$ and 
$A=\frac{{G_{\chic F}}^2}{2 {\pi}^3}({g_V}^2+3{g_A}^2)$. $G_{\chic F}$ is 
the Fermi coupling constant, and $P$ ($Q$) is a factor of survival 
probability ($<1$) for neutrinos (antineutrinos). In this work, we assume 
the energy spectrum of each neutrino species, electron and positron is 
approximated by a Maxwell-Boltzmann distribution, following Bernstein 
\cite{Bernstein} and Enqvist \cite{Enqvist}.

We changed to the variable $y=\Delta m/T$ and the solution is function 
of the temperature, the integrating factor now becomes 
\begin{equation}
I(y,y^{'})= Exp \left( {-\int_{y^{'}}^y 
\left[\frac{dt^{''}}{dy^{''}}\right]\lambda_2(y{''})dy{''}} \right).
\end{equation}
To evaluate the integrating factor we assume the neutrino-driven wind 
model, the radius of an outflowing mass element is related to time $t$ by 
$r=r_o exp\left\{ {(t-t_o)/t_{dyn}} \right\}$ this implies an outflow 
velocity proporcional to the radius, $v=r/t_{dyn}$ \cite{Meyer}. In the 
model, $\rho \, \alpha \,  r^{-3}$ and $T \, \alpha \, r^{-1}$. We set 
$T_n = T_p  = T_e= T_{e^+}=T$, but we assume that the neutrino and 
antineutrino temperatures remain nearly constant. We finally compute 
the integrals in the solution for a range of $y$ values, we obtain the 
curve for $Y_e(y)$ and $P$. Numerical supernova neutrino transport 
calculations show that $T_{\overline{\nu}_e}= 5.1$ Mev, $T_{\nu_e} = 4.5$ 
Mev and we set for $\tau_{dyn} =0.3$ s. Neutrino oscillations add a new 
complication to the diagnostic of supernova neutrinos, the factor $P$ 
will be the survival probability of the electron neutrinos and 
antineutrinos.
\section{Conclusions}
In Figures (a) and (b) show that difference betwenn $Ye$ (doted line) and 
$Yeq$ (solid line) is very little. That is, the integral in the equation (4) 
not contribute significantly to $Ye$. Canculations made for $\tau_{dyn}$ 
between $0.01\, s$ and $1\, s$ confirm the last.
\begin{acknowledgments}
This work was supported in part by the {\it Programa de Apoyo a
Proyectos de Investigación e Inovación Tecnológica} 
({\bf PAPIIT}) de la {\bf DGAPA-UNAM} {\it No. de Proyecto}: 
{\sc IN109001} and in part by {\bf CONACYT} under proyect No I37307-E.
\end{acknowledgments}
\end{document}